# pyTransRate: reference-free quality assessment of *de novo* transcriptome assemblies with current aligners and quantifiers

Matthew D. MacManes

Department of Molecular, Cellular and Biomedical Sciences, University of New Hampshire, Durham, NH 03824, USA

## Abstract

Summary: The original software TransRate scores a *de novo* transcriptome assembly by mapping the reads it was built from back to it, with no reference required. Its last release (1.0.3) depends on Ruby 2.2 and on aligner and quantifier versions that can no longer be installed on most systems. pyTransRate is a Python reimplementation of the TransRate method built on snap-aligner 2.0.5 and salmon 2.7.0. In 5,031 paired comparisons (39 assembly-level metrics on each of 129 public assemblies), it reproduces the original sequence metrics and read counts exactly and ranks assemblies consistently with 1.0.3 (Spearman $\varrho = 0.94$ for assembly score, 0.98 for good-mapping rate), although its read-based scores are not numerically interchangeable with the original. It also reads gzip- and bgzip-compressed input directly.

Availability and Implementation: pyTransRate is implemented in Python 3 (≥3.9) and is freely available under the MIT license at https://github.com/macmanes-lab/pytransrate. It runs on Linux and macOS; snap-aligner and salmon are installed from Bioconda.

Contact: matthew.macmanes@unh.edu

Supplementary information: Supplementary data are available online. Supplementary Table S1 lists the assembly and read accessions for every assembly analyzed; Supplementary Table S2 gives per-assembly values from both implementations.

## 1 Introduction

Every year, *de novo* transcriptome assemblies are built for thousands of model and non-model organisms, and choosing among assemblies, or among contigs within one, requires a measure of quality that does not depend on a reference genome. TransRate (Smith-Unna et al., 2016) provides a robust and well-vetted method for doing this. It maps the input reads back to the assembly, assigns each fragment to a single contig, and summarizes how well each contig is supported by four components: per-base agreement with the reads (sCnuc), breadth of coverage (sCcov), consistency of read-pair placement (sCord) and uniformity of coverage (sCseg). Their product is the contig score. The assembly score is the geometric mean of contig scores multiplied by the proportion of fragments that map consistently (the good-mapping rate). TransRate is widely used for assembly comparison and filtering, including within the Oyster River Protocol (MacManes, 2018), alongside other tools such as BUSCO (Manni et al., 2021).

The reference implementation, TransRate 1.0.3, can no longer be installed or run (e.g., https://github.com/blahah/transrate/issues/240, https://github.com/blahah/transrate/issues/253) on most current systems. It is pinned to Ruby 2.2.0 (end of life since 2018), is distributed as an x86_64 Linux bundle, and depends on snap-aligner 1.0dev.96, salmon 0.8.2 which are no longer maintained. Its automatic dependency installer fetches binaries from URLs that no longer resolve.

pyTransRate is a Python reimplementation of the TransRate method against current dependencies. It keeps the original formulas, output columns and column order. Because the updated aligner and quantifier change the inputs to those formulas, pyTransRate does not reproduce the numerical scores of 1.0.3, and the software was renamed so that the two sets of numbers are not compared by accident. Here we describe the implementation and quantify how its scores relate to those of TransRate version 1.0.3 on 129 public assemblies scored by both implementations.

# 2 Implementation

## 2.1 Pipeline

pyTransRate takes an assembly in FASTA format and paired-end Illumina sequencing reads. Sequence metrics (contig lengths, Nx, GC content, longest open reading frame) are computed. Reads are aligned with snap-aligner 2.0.5 (Zaharia et al., 2011) allowing multiple alignments per fragment, contig abundances are estimated with salmon 2.7.0 (Patro et al., 2017), and each fragment is assigned to one contig. Per-contig coverage, edit distance and pairing statistics are computed directly from the BAM stream using pysam and htslib (Bonfield et al., 2021). Outputs are contigs.csv (per-contig metrics and scores), assemblies.csv (assembly-level metrics, assembly score and optimal score) and the score-optimization curve used to choose a contig-score cutoff. Column order is unchanged from 1.0.3. A typical run is

```
pytransrate -a assembly.fa --left R1.fq.gz --right R2.fq.gz -t 16 -o results
```

## 2.2 Changes that alter scores

Three changes to the inputs of the TransRate formulas mean that scores differ from 1.0.3. First, snap-aligner 2.0 soft-clips reads by default. The original bam-read helper advanced its reference position across soft clips, which shifted the coverage of every clipped read; pyTransRate follows SAM specification, and its per-base coverage agrees with samtools depth (Danecek et al., 2021). Second, salmon 2.7.0 no longer writes the posterior-sampled alignments (postSample.bam) that 1.0.3 used to assign fragments. pyTransRate therefore assigns each fragment deterministically to its maximum a posteriori contig, using salmon abundances as the prior and alignment edit distance as the likelihood, with ties broken by prior and then contig name. Third, default snap-aligner settings (-H 300000 -D 5 -om 5 -omax 10) were reduced (-H 4000 -D 2 -om 2 -omax 10) because the original settings cause snap-aligner 2.0.5 to fail on real assemblies.

## 2.3 Changes impacting usage

Several additions address failures that appear only on large assemblies. Assembly and read files may be gzip- or bgzip-compressed in any combination; compression is detected from the file contents rather than the name, and reads are passed to snap-aligner compressed. Contig names are taken up to the first space, as snap-aligner and salmon do, so NCBI TSA and ENA headers containing | are handled correctly. Scoring is parallelized across processes, and the number of processes is capped at what fits within the memory available from the Slurm allocation, the cgroup limit or --max-memory flag. An interrupted run can be restarted with the same command: the snap-aligner index, BAM and salmon output are reused after checks that each is complete. Implementation details, including profiling and a list of changes by version, are in the repository changelog.

## 3 Comparison with TransRate 1.0.3

We scored 129 public assemblies with both TransRate 1.0.3 and pyTransRate, using the same assembly and read files for each. They represent 103 eukaryotic species: 15 were assembled with the Oyster River Protocol from SRA, ENA or DDBJ runs and 114 were obtained from the NCBI Transcriptome Shotgun Assembly (TSA) archive. They span Metazoa (95), Viridiplantae (25), Fungi (4) and five other eukaryotic groups, and range from 5,050 to 625,091 contigs and from 6.6 to 467 million read pairs. All 1.0.3 runs used decompressed reads because a bug in the original 1.0.3 code prevented proper read of compressed files. For each assembly we compared 39 assembly-level metrics, 18 derived from the sequences alone and 21 derived from read mapping, for a total of 5,031 direct comparisons. Assembly and read accessions are listed in Supplementary Table S1 and per-assembly values in Supplementary Table S2.

All 2,322 sequence-metric comparisons (contig count, length distribution, Nx values, GC content, N content, number of contigs with an ORF and mean ORF coverage) were identical. The number of read pairs counted was also identical in all 129 assemblies, so every difference reported below arises from mapping, fragment assignment rather than from the input.

Assembly scores from the two implementations were strongly correlated (Fig. 1A; Spearman $\varrho = 0.94$, Kendall $\tau = 0.82$). pyTransRate scored 101 of 129 assemblies higher; the median paired difference was small (+0.013; interquartile range +0.003 to +0.041) but not zero (Wilcoxon signed-rank $p = 2.7 \times 10^{-8}$), and individual assemblies moved by as much as −0.151 and +0.129. Scores differed by 0.02 or less for 70 of 129 assemblies (54%) and by 0.05 or less for 97 (75%). The optimal score, obtained after discarding contigs below the selected cutoff, agreed more closely (Fig. 1C; $\varrho = 0.98$, $\tau = 0.91$, median difference 0.000; 92 assemblies within 0.02), with only a slight asymmetry in the paired differences ($p = 0.04$). The cutoff chosen to reach it was less consistent between implementations ($\varrho = 0.72$).

Good-mapping rates were highly concordant (Fig. 1B; $\varrho = 0.98$, $\tau = 0.93$) and were slightly higher in pyTransRate for 128 of 129 assemblies (median +0.015; interquartile range +0.008 to +0.024; $p = 3.4 \times 10^{-22}$). They differed by 0.02 or less for 86 assemblies (67%), and three assemblies gained more than 0.10 (maximum +0.215). The proportion of fragments mapped rose in 128 of 129 (median +0.022). Because read counts were identical, this shift reflects more reads aligning under snap-aligner 2.0.5 and its settings rather than any difference in input.

The contig-level components that make up the score also agreed closely in rank ($\varrho = 0.92$–$0.99$; Table 1) but shifted in consistent directions. The proportion of bases without read coverage rose (median +0.014), as did the proportion of contigs flagged as segmented (+0.017), while the proportions of contigs with no coverage (−0.021) and with low coverage (−0.015) fell. These opposing shifts partly offset one another in the contig score, so the assembly score changes less than some of its components. Ranking of assemblies built from the same reads is therefore largely preserved, but scores from the two implementations should not be mixed in one analysis.

Table 1. Read-based assembly-level metrics for 129 assemblies scored by TransRate 1.0.3 and pyTransRate. Values are medians (interquartile range); Δ is the median of per-assembly differences (pyTransRate minus 1.0.3); $\varrho$ is Spearman's rank correlation. Proportions of contigs are of all contigs in the assembly.

| Metric | TransRate 1.0.3 | pyTransRate 2.2.2 | Median Δ | $\varrho$ |
|---|---|---|---|---|
| Assembly score | 0.227 (0.124–0.343) | 0.255 (0.147–0.344) | +0.013 | 0.94 |
| Optimal score | 0.381 (0.221–0.468) | 0.385 (0.230–0.456) | 0.000 | 0.98 |
| Good-mapping rate | 0.740 (0.507–0.825) | 0.756 (0.519–0.857) | +0.015 | 0.98 |
| Fragments mapped | 0.847 (0.692–0.919) | 0.876 (0.742–0.937) | +0.022 | 0.97 |
| Bases uncovered | 0.076 (0.029–0.145) | 0.108 (0.049–0.180) | +0.014 | 0.93 |
| Contigs uncovered | 0.052 (0.016–0.094) | 0.023 (0.004–0.050) | −0.021 | 0.94 |
| Contigs low-coverage | 0.359 (0.194–0.497) | 0.334 (0.179–0.475) | −0.015 | 0.99 |
| Contigs segmented | 0.096 (0.063–0.122) | 0.119 (0.074–0.147) | +0.017 | 0.92 |

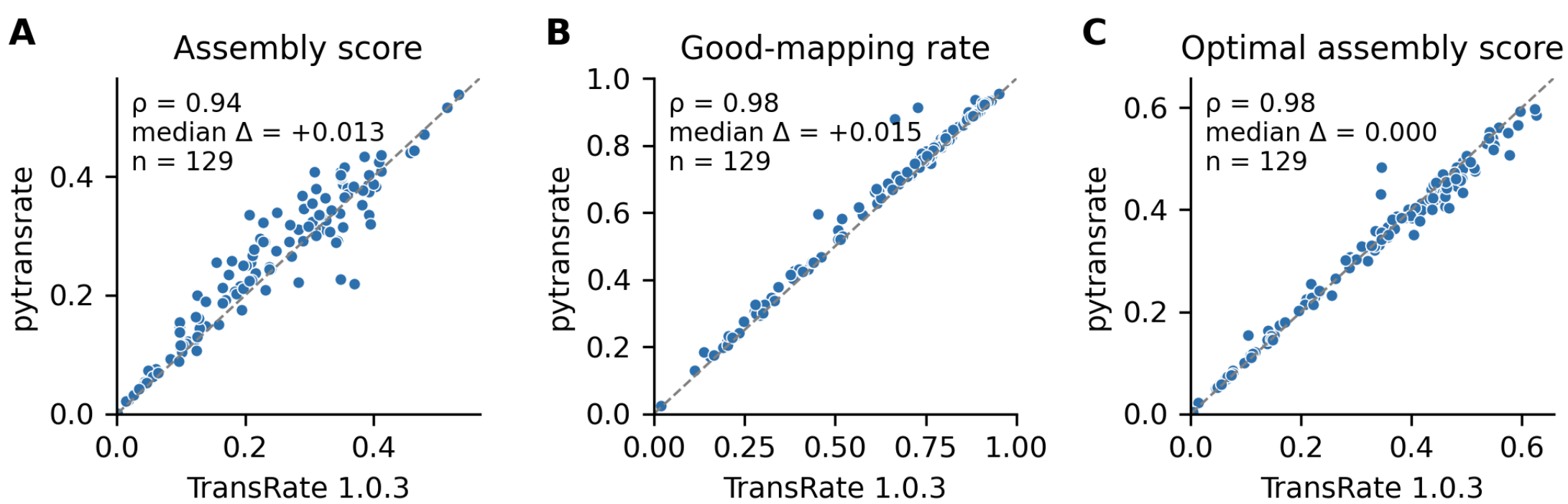

Fig. 1. Agreement between TransRate 1.0.3 and pyTransRate 2.2.2 for 129 assemblies scored by both implementations. (A) Assembly score, (B) good-mapping rate and (C) optimal assembly score. Each point is one assembly; the dashed line marks equality. $\varrho$, Spearman's rank correlation; Δ, median paired difference (pytransrate minus 1.0.3).

## 4 Discussion and Conclusions

pyTransRate makes the TransRate method usable again. It reproduces the original sequence metrics exactly and ranks assemblies consistently with version 1.0.3, but its read-based scores are not interchangeable with 1.0.3 values. Scores remain comparable only between assemblies built from the same reads, as they were in the original method.

Of note, the reference-based metrics from TransRate version 1.0.3 are not implemented.

The method, scoring model and original software are the work of Smith-Unna et al. (2016), and users of pyTransRate should cite that paper in addition to this one.

## Conflict of Interest

Conflict of interest: none declared.

## AI Statement

For full disclosure, Claude was used to port over the code from Ruby to Python and was used to identify and fix bugs in the subsequent code. Code was reviewed and tested by MacManes. Further, AI was used to make some of the passages in this document clearer.

Experiments were conceived of, designed, and monitored by MacManes. All results and claims have been validated by the author. The code is freely available on GitHub and is fully executable. All data on which the claims in this paper are made are included in the supplemental.

## Acknowledgements

We thank the authors of TransRate for releasing their software under the MIT license.

## Data availability

The pyTransRate source code is available at https://github.com/macmanes-lab/pytransrate. The assemblies analyzed are available from NCBI TSA, and the reads from the SRA, ENA and DDBJ, under the accessions listed in Supplementary Table S1. Per-assembly metrics from both implementations (Supplementary Table S2) and the scripts used to produce Table 1 and Figure 1 are archived on GitHub.